\documentclass{iopart}

\usepackage{graphicx}
\usepackage{epsfig}

\begin{document}

\title{Three particle decays of light nuclei resonances}

\author{R~\'Alvarez-Rodr\'{\i}guez$^1$, A~S~Jensen$^2$, E~Garrido$^3$,
  D~V~Fedorov$^2$}

\address{ $^1$ Departamento de F\'{\i}sica e Instalaciones, ETS
  Arquitectura, Universidad Polit\'ecnica de Madrid, E-28040 Madrid,
  Spain\\ $^2$ Department of Physics and Astronomy, University of
  Aarhus DK-8000 Aarhus C, Denmark\\ $^3$ Instituto de Estructura de
  la Materia, Consejo Superior de Investigaciones Cient\'{\i}ficas
  E-28006 Madrid, Spain}

\ead{raquel.alvarez@upm.es}

\begin{abstract}
We have studied the three-particle decay of $^{12}$C, $^9$Be and
$^6$Be resonances. These nuclei have been described as three-body
systems by means of the complex scaled hyperspherical adiabatic
expansion method. The short-distance part of the wave-function is
responsible for the energies whereas the information related to the
observable decay properties is contained at large distances, which must
be computed accurately. As an illustration we show the results for the
angular distribution of $^9$Be and $^6$Be resonances.
\end{abstract}


\submitto{\PS}


\section{Introduction.}

In Quantum Mechanics the decay of two fragments is well determined by
energy and momentum conservation laws. But the decay of three
particles is a much more complicated issue. The energies of
the decaying fragments are not fixed. The energy distribution of the
three-body final state after the decay is an observable, and can be
used to understand the structure of the initial state and the decay
mechanism itself. On the other hand, the intermediate path connecting
the initial and final states is not an observable. The only way to
extract information about the decay mechanism is to try to understand
the measurable final state by means of theoretical models. The decay
is usually interpreted as either sequential via an intermediate
configuration or direct to the continuum. The interpretations are
often used to derive reaction rates for the inverse process in
astrophysical environments. It is therefore important to have a
reliable interpretation of the data.

We have focused our attention on three-body decaying nuclei involved
in stellar nucleosynthesis reactions, $^{12}$C, $^9$Be and $^6$Be. For
all three there are experimental data available that help us to test
the validity of our theoretical model.

\section{Theoretical framework.}

The three-body decaying nuclei are described as three-body systems
within the complex-scaled hyperspherical adiabatic expansion method
\cite{nie01}. According to this method, The angular part of the
Hamiltonian is first solved keeping fixed the value of the hyperradius
$\rho$. Its eigenvalues serve as effective potentials while the
eigenfunctions, $\Phi_{nJM}$ are used as a basis to expand the total
wave-function $\Psi^{JM} = \frac{1}{\rho^{5/2}}\sum_n f_n (\rho)
\Phi_{nJM} (\rho,\Omega)\;.  $ The $\rho$-dependent expansion
coefficients, $f_n (\rho)$, are the hyperradial wave functions
obtained from the coupled set of hyperradial equations.

$^{12}$C is described as three $\alpha$-particles, $^9$Be as two
$\alpha$-particles and one neutron, and $^6$Be as one
$\alpha$-particle and two protons. Our Hamiltonian contains
short-range and Coulomb potential (between charged particles). We have
considered $\alpha-\alpha$ potential from \cite{ali}, $\alpha$-nucleon
from \cite{cobis} and nucleon-nucleon from \cite{garrido}. These
potentials have been built in order to reproduce the two-body
scattering data. On top of these interactions we include a
structureless three-body potential of the form $V_{3b} =
S\exp(-\rho^2/b^2)$, fitted to reproduce the resonance energies. This
potential is included because at short distances the structure of
these nuclei is not necessarily of three-body character. The complex
scaling method helps us to treat the resonances as if they were bound
states.

The many-body initial state resonance evolves into three clusters at
large distances. The total angular momentum and parity $J^\pi$ is
conserved in the process. This symmetry imposes constraints on the
resulting momentum distributions. The energy distribution is the
probability for finding a given particle at a given energy. It can be
measured experimentally and is the only information that allows us to
study the decay path, that can be either sequential or direct or a
mixture. The information about the energy distributions of the
fragments after the decay is contained in the large-distance part of
the wave-function, which must be accurately computed. The single
particle probability distributions are obtained after Monte Carlo
integration of the absolute square of the large-distance
wave-function.

\section{Results.}

\subsection{$^{12}$C}

Within this theoretical framework we have extensively studied the
decay of the low-lying $^{12}$C resonances \cite{alv08}. Our results
have been compared to recent experimental data \cite{oli} with a high
level of agreement. Moreover, our suggestion to change the previously
assigned spin and parity of the 13.35~MeV state \cite{alvdres} from
$2^-$ to $4^-$ has been supported by the experimental community
\cite{oli,freer}.

\subsection{$^9$Be}

\begin{figure}
\vspace*{0.1cm}
\epsfig{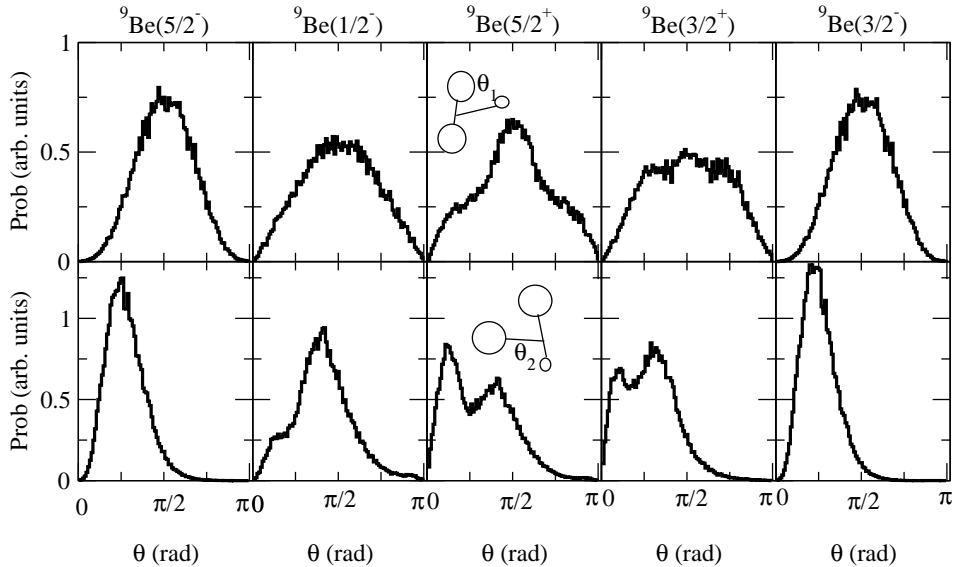}
\vspace*{0.3cm}
\caption{The angular distributions of the directions between two
  particles and their centre-of-mass and the third particle for the
  $5/2^-$, $1/2^-$, $5/2^+$, $3/2^+$ and $3/2^-$ resonances of $^9$Be
  as labelled in the figure. The distributions have been computed for
  the two possible angles that can be defined in our three-body
  system.}
\label{fig9}
\end{figure}

We have studied the five lowest resonances of $^9$Be \cite{alv10} and
compared them to the experimental data from
\cite{fulton}. Fig.~\ref{fig9} shows the angular distributions of
these resonances, {\it i.e.} the probability for finding one of the
decaying particles in a certain direction with respect to the
direction formed by the other two. In all the cases we have removed
the sequential decay via $^8$Be($0^+$). This kind of plot contains
information about the angular momentum of the first particle relative
to the centre-of-mass of the other two. We observe that the angular
distribution patterns are different for different $J^\pi$
states. These features are clearly distinguishable, demonstrating that
these observables can be used to determine the large-distance
structure of these resonances. The initial state can still be
determined only through the theoretical information about the
dynamical evolution of the resonances.

\subsection{$^6$Be}

\begin{figure}
\vspace*{0.1cm}
\epsfig{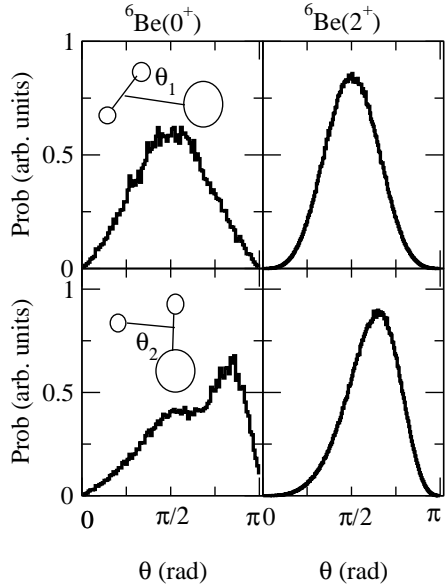}
\vspace*{0.3cm}
\caption{Same as in fig.~\ref{fig9}, but for the $0^+$ and $2^+$
  resonances of $^6$Be.}
\label{fig6}
\end{figure}

The $^6$Be is an unbound nucleus which has only two low-lying
resonances, $0^+$ and $2^+$. Its decay has been recently measured
\cite{papka} and compared to theoretical predictions given by our
formalism. We have not found any signatures of sequential decay of
these resonances via intermediate two-body states. We show in
fig.~\ref{fig6} the angular distributions for the two resonances of
$^6$Be. The upper panels can be compared directly to the experimental
data from \cite{papka}. They show that the $\alpha$-particle prefers
to come out perpendicular to the direction formed by the two protons.

\section{Summary and Conclusions.}

We have applied a general method to compute the momentum distributions
of three-body decaying light-nuclei resonances. We have
conjectured that the energy distributions of the decay fragments are
insensitive to the initial many-body structure. The energy
distributions are then determined by the energy and three-body
resonance structure as obtained in a three-body cluster model. These
momentum distributions are determined by the coordinate space
wave-functions at large distances, which much be computed with a high
accuracy.

The method has been applied previously to the study of the decay of
$^{12}$C resonances with great success. In this contribution we have
shown the angular distribution of the low-lying $^9$Be and $^6$Be
resonances decaying into $\alpha+\alpha+n$ and $\alpha+p+p$
respectively. Our distributions are open to experimental
tests.

\ack This work has been partly supported by funds provided by MICINN
(Spain) under the contracts FIS2008-01301 and FPA2010-17142 and the
Spanish Consolider-Ingenio programme CPAN (Programme
No. CSD2007-00042).

\end{document}